\documentstyle[amssymb,preprint,aps]{revtex}
\begin{document}
\title{Three-Fluid Description of the Sympathetic Cooling of a Boson-Fermion Mixture}
\author{M. Wouters, J. Tempere$^{\ast }$, J. T. Devreese$^{\ast \ast }$}
\address{Departement Natuurkunde, Universiteit Antwerpen, Universiteitsplein 1, B2610%
\\
Antwerpen, Belgium.}
\date{May 2nd, 2002.}
\maketitle

\begin{abstract}
We present a model for sympathetic cooling of a mixture of fermionic and
bosonic atomic gases in harmonic traps, based on a three-fluid description.
The model confirms the experimentally observed cooling limit of about 0.2 $%
T_{F}$ when only bosons are pumped. We propose sequential cooling -- first
pumping of bosons and afterwards fermions -- as a way to obtain lower
temperatures. For this scheme, our model predicts that temperatures less
than $0.1$ $T_{F}$ can be reached.
\end{abstract}


\section{Introduction}

\tighten Recent attempts to cool down a trapped gas of fermionic atoms, such
as $^{40}$K \cite{demarcoSCI285} or $^{6}$Li \cite
{truscottSCI291,schreckPRL87}, have reached temperatures below the Fermi
temperature. Degenerate{\bf \ }atomic Fermi gases offer the intriguing
possibility to create a paired fermion state such as a
Bardeen-Cooper-Schrieffer phase in a novel and highly controllable system.
However, until{\bf \ }now the temperatures obtained in the experiments with
ultracold fermionic atoms are not low enough to observe such phenomena. In
this paper, we propose a straightforward description of the cooling
mechanism, discuss its inherent limitations, and we investigate strategies
to alter the cooling procedure in order to achieve lower temperatures.

The cooling process used to create Bose-Einstein condensates in trapped
bosonic gases is evaporative cooling, whereby the most energetic atoms are
removed from the gas, leaving the remnant colder after rethermalization.
However, due to the antisymmetrization requirement of the fermionic wave
function, identical fermionic atoms cannot undergo the $s$-wave collisions
necessary for rethermalization, and thus they cannot be directly cooled
using evaporative cooling. This problem was circumvented by simultaneously
trapping two atom species or two different hyperfine spin states of a given
atom species. In \cite{truscottSCI291}, $^{6}$Li is trapped together with
its bosonic isotope $^{7}$Li. Whereas the $s$-wave scattering length of $%
^{6} $Li is zero by symmetry, the $s$-wave scattering length for a collision
between a $^{6}$Li and a $^{7}$Li atom is 2.2 nm, which is enough to
thermalize both species together. This version of evaporative cooling is
denominated `sympathetic cooling'.

We describe the sympathetic cooling process, using the three-fluid model
presented in Sec. II, as a sequence of evaporation and rethermalization
steps so that in the evaporation steps the energetic atoms are removed from
the mixture and in the rethermalization steps the new (lower) equilibrium
temperature is reached, as explained in Sec. III. The results obtained with
this model are discussed in Sec. IV, and an improvement on the current
sympathetic cooling process is proposed to reach lower temperatures ($<0.1$ $%
T/T_{F}$).

\section{Three-fluid model}

The trapped system which will be subjected to sympathetic cooling consists
of $N_{b}$ bosons with mass $m_{b}$ and scattering length $a_{b}$; and $%
N_{f} $ fermions with mass $m_{f}$, all at initial temperature $T$ in a
parabolic trap with trapping frequency $\omega $ and characteristic length%
{\em \ }$a_{\text{HO}_{b,f}}=\sqrt{\hbar /\left( m_{b,f}\omega \right) }$.
The scattering length for a collision between a boson and a fermion will be
denoted by $a_{f}$. The three-fluid model \cite{amorusoEPJD4} distinguishes
the fermion gas, the Bose-Einstein condensed (BEC) bosons and the
non-condensed or `thermal' bosons. The number of BEC bosons is $%
N_{c}=N_{b}(1-(T/T_{C})^{3})$ for $T<T_{C}$ and zero otherwise, where \cite
{giorginiPRA54} 
\begin{equation}
T_{C}=\frac{\hbar \omega }{k_{B}}\sqrt[3]{\frac{N_{b}}{\zeta \left( 3\right) 
}}\left( 1-0.73N_{b}^{-1/3}-1.33\frac{a_{b}}{a_{\text{HO}_{b}}}%
N_{b}^{1/6}\right) .
\end{equation}
The number of thermal bosons is $N_{t}=N_{b}-N_{c}$. We use the notation 
\begin{equation}
g=\frac{4\pi \hbar ^{2}a_{b}}{m_{b}}\text{ and }f=\frac{4\pi \hbar ^{2}a_{f}%
}{m_{r}}
\end{equation}
where $m_{r}=2m_{b}m_{f}/\left( m_{b}+m_{f}\right) $ is twice the reduced
mass of a fermion and a boson. $n_{c}({\bf r})$ is the density of BEC
bosons, $n_{t}({\bf r})$ is the density of thermal bosons, and $n_{f}({\bf r}%
)$ is the density of fermions. In the three-fluid model, the BEC bosons, the
thermal bosons and the fermions are described as ideal gases in effective
one-body potentials. The effective one-body potential for the BEC bosons ($%
V_{c}$), for the thermal bosons ($V_{t}$) and for the fermions ($V_{f}$) are
given by: 
\begin{eqnarray}
V_{c}({\bf r}) &=&\frac{m_{b}\omega ^{2}}{2}r^{2}+gn_{c}\left( {\bf r}%
\right) +2gn_{t}({\bf r})+fn_{f}({\bf r})  \label{Vc} \\
V_{t}({\bf r}) &=&\frac{m_{b}\omega ^{2}}{2}r^{2}+2gn_{c}({\bf r})+2gn_{t}(%
{\bf r})+fn_{f}({\bf r})  \label{Vt} \\
V_{f}({\bf r}) &=&\frac{m_{f}\omega ^{2}}{2}r^{2}+fn_{c}({\bf r})+fn_{t}(%
{\bf r})  \label{Vf}
\end{eqnarray}
These potentials determine the densities 
\begin{eqnarray}
n_{c}({\bf r}) &=&\frac{1}{g}\left[ \mu _{b}-\frac{m_{b}\omega ^{2}}{2}%
r^{2}-2gn_{t}({\bf r})-fn_{f}({\bf r})\right]  \label{nc} \\
n_{t}({\bf r}) &=&\int \frac{d^{3}{\bf k}}{(2\pi )^{3}}\frac{1}{{\bf \exp }%
\left[ \beta \left( 
%TCIMACRO{\dfrac{\hbar ^{2}k^{2}}{2m_{b}}}%
%BeginExpansion
{\displaystyle{\hbar ^{2}k^{2} \over 2m_{b}}}%
%EndExpansion
+V_{t}({\bf r})-\mu _{b}\right) \right] -1}  \label{nt} \\
n_{f}({\bf r}) &=&\int \frac{d^{3}{\bf k}}{(2\pi )^{3}}\frac{1}{1+{\bf \exp }%
\left[ \beta \left( 
%TCIMACRO{\dfrac{\hbar ^{2}k^{2}}{2m_{f}}}%
%BeginExpansion
{\displaystyle{\hbar ^{2}k^{2} \over 2m_{f}}}%
%EndExpansion
+V_{f}({\bf r})-\mu _{f}\right) \right] }  \label{nf}
\end{eqnarray}
In these expressions, $\beta =1/(k_{B}T)$ and $\mu _{f},\mu _{b}$ is the
chemical potential of the fermion gas and boson gas respectively. The set of
equations (\ref{Vc})-(\ref{Vf}) and (\ref{nc})-(\ref{nf}) form a
self-consistent description of the three fluids, in which the chemical
potentials are set by fixing the total number of fermions and the total
number of bosons; $N_{b}=\int d^{3}{\bf r}(n_{c}({\bf r})+n_{t}({\bf r}))$
and $N_{f}=\int d^{3}{\bf r}$ $n_{f}({\bf r}).$ Note that the interaction
between the bosonic atoms, and the interaction between the bosonic and
fermionic atoms, is taken into account on the level of the mean-field
approximation.

The total energy of the BEC bosons ($E_{c}$), of the thermal bosons ($E_{t}$%
) and the fermions ($E_{f}$) is given by 
\begin{eqnarray}
E_{c} &=&\int d^{3}{\bf r}\,\left[ \frac{m_{b}\omega ^{2}}{2}%
r^{2}n_{c}\left( {\bf r}\right) +\frac{1}{2}g\left[ n_{c}\left( {\bf r}%
\right) \right] ^{2}\right]  \label{Ec} \\
E_{t} &=&\int d^{3}{\bf r}\,\left[ \int \frac{d^{3}{\bf k}}{(2\pi )^{3}}%
\frac{\frac{m_{b}\omega ^{2}}{2}r^{2}+\frac{m_{b}}{2}k^{2}}{{\bf \exp }\left[
\beta \left( 
%TCIMACRO{\dfrac{\hbar ^{2}k^{2}}{2m}}%
%BeginExpansion
{\displaystyle{\hbar ^{2}k^{2} \over 2m}}%
%EndExpansion
+V_{t}({\bf r})-\mu _{b}\right) \right] -1}+g\left[ n_{t}\left( {\bf r}%
\right) \right] ^{2}+2gn_{t}\left( {\bf r}\right) n_{c}\left( {\bf r}\right) %
\right]  \label{Et} \\
E_{f} &=&\int d^{3}{\bf r}\,\left[ \int \frac{d^{3}{\bf k}}{(2\pi )^{3}}%
\frac{\frac{m_{f}\omega ^{2}}{2}r^{2}+\frac{m_{f}}{2}k^{2}}{{\bf \exp }\left[
\beta \left( 
%TCIMACRO{\dfrac{\hbar ^{2}k^{2}}{2m_{f}}}%
%BeginExpansion
{\displaystyle{\hbar ^{2}k^{2} \over 2m_{f}}}%
%EndExpansion
+V_{f}({\bf r})-\mu _{f}\right) \right] +1}+fn_{f}\left( {\bf r}\right)
\left( n_{c}\left( {\bf r}\right) +n_{t}\left( {\bf r}\right) \right) \right]
.  \label{Ebf}
\end{eqnarray}
From the above set of equations (\ref{Vc})-(\ref{Ebf}), the temperature, and
the chemical potential of the bosons and the fermions can be calculated if
the total numbers of bosons and fermions, and their total energy are given.
Note that the Bose-Einstein condensate is treated in the Thomas-Fermi
approximation as if it were at zero temperature. For the thermal bosons we
have experimented using the Maxwell-Boltzman distribution instead of a
Bose-Einstein distribution in the expressions for the density (\ref{nt}) and
the energy (\ref{Et}). In the experiment, this approximation is routinely
performed in order to determine the temperature of a trapped atomic gas by
fitting the Maxwell-Boltzman distribution to the tail of the measured
velocity distribution or density profile. Also in our calculations, we found
that the Maxwell-Boltzman distribution provides an accurate description of
the thermal bosons.

\section{Sympathetic cooling}

We describe the cooling process as a two-step process. The first step is the 
{\it evaporative stage}, whereby either just the bosons can be `evaporated',
or just the fermions, or both bosons and fermions. The effect of the
evaporative stage is truncating the radial distribution by removing the
outer atoms. The energy of these atoms is higher than the average energy of
the mixture and so this removal allows us to obtain a lower temperature. The
second step is the {\it rethermalization stage}, where the collisions
between the bosons and between bosons and fermions bring the truncated
(non-equilibrium) distribution obtained in the evaporative stage into a new
equilibrium distribution. This new equilibrium distribution will be at a
lower temperature if the atoms removed have more energy than the average
atom energy. We will describe the sympathetic cooling as a discrete
succession of evaporation and rethermalization stages. In reality, the
evaporation and rethermalization goes on simultaneously. However, previous
studies have shown that the discrete treatment of the cooling process
provides an accurate description \cite{davisAPB60}.

The {\it evaporation }is realized by a radiofrequent (or microwave) field
which induces a transition between the trapped hyperfine state of a selected
atom species and an untrapped hyperfine state. The frequency suited \ for
this transition depends on the distance from the atom to the center of the
trap, due to the Zeeman effect of the spatially varying magnetic trapping
field. By tuning the frequency, atoms that are beyond a cut-off distance $R$
from the center of the trap are removed. If the bosons are evaporated in
this way, then the number of bosons which are removed from the mixture is 
\begin{equation}
\Delta N_{t}=%
%TCIMACRO{\dint}%
%BeginExpansion
\displaystyle\int %
%EndExpansion
\limits_{r>R}d^{3}{\bf r\,}n_{t}\left( {\bf r}\right) ,
\end{equation}
and the total energy removed in the evaporation is 
\begin{equation}
\Delta E_{t}=%
%TCIMACRO{\dint}%
%BeginExpansion
\displaystyle\int %
%EndExpansion
\limits_{r>R}d^{3}{\bf r}\,\left[ \int \frac{d^{3}{\bf k}}{(2\pi )^{3}}\frac{%
\frac{m_{b}\omega ^{2}}{2}r^{2}+\frac{m_{b}}{2}k^{2}}{{\bf \exp }\left[
\beta \left( 
%TCIMACRO{\dfrac{\hbar ^{2}k^{2}}{2m}}%
%BeginExpansion
{\displaystyle{\hbar ^{2}k^{2} \over 2m}}%
%EndExpansion
+V_{t}({\bf r})-\mu _{b}\right) \right] -1}+g\left[ n_{t}\left( {\bf r}%
\right) \right] ^{2}+\,fn_{f}\left( r\right) n_{t}\left( r\right) \right] .
\end{equation}
If fermionic atoms are removed from the mixture, then the corresponding
number of removed fermions $\Delta N_{f}$ and the total evaporated energy of
the fermion gas $\Delta E_{f}$ are found from expressions analogous to those
above. Note that evaporating the Bose-Einstein condensed bosons will not
lead to cooling. The Bose-Einstein condensed bosons are always the
lowest-energetic bosons in the mixtures - removing them would only lead to
heating of the system. Thus it is important that the cut-off distance $R_{C}$
is larger than the Thomas-Fermi radius of the condensate.

The evaporative stage results in a new total number of bosons $N_{b}^{\prime
}=N_{b}-\Delta N_{b}$, and a new total energy $E_{\text{total}}^{\prime
}=E_{c}+E_{t}+E_{f}-\Delta E_{t}$, if only bosons are removed from the trap.
If also fermions are evaporated, $\Delta E_{f}$ has to be subtracted too,
and $N_{f}^{\prime }=N_{f}-\Delta N_{f}$ is the new number of fermions.
Using the equations (\ref{Vc})-(\ref{Ebf}), the new system parameters $%
N_{f}^{\prime },N_{b}^{\prime },E_{\text{total}}^{\prime }$ are expressed as
a function of the new temperature $T^{\prime }$ and the new chemical
potentials $\mu _{b}^{\prime }$ and $\mu _{f}^{\prime }$. The new
temperature and chemical potentials are then solved from these equations.
After this rethermalization stage, a new evaporative stage can be initiated.

Losses due to three-body collisions lead to heating, since the three-body
collisions will be prominent in the condensate which harbors the coldest
atoms. Such losses introduce another depletion of the number of atoms. For
the bosons, 
\begin{eqnarray}
\Delta N_{b}^{\text{heat}} &=&K_{3}t%
%TCIMACRO{\dint }%
%BeginExpansion
\displaystyle\int %
%EndExpansion
d{\bf r}\text{ }[n_{c}({\bf r})]^{3}, \\
\Delta E_{b}^{\text{heat}} &=&K_{3}t%
%TCIMACRO{\dint }%
%BeginExpansion
\displaystyle\int %
%EndExpansion
d{\bf r}\text{ }[n_{c}({\bf r})]^{3}\left( \frac{m_{b}\omega ^{2}}{2}%
r^{2}+g\,n_{c}\left( {\bf r}\right) +2g\,n_{t}\left( {\bf r}\right) +f{\bf \,%
}n_{f}\left( r\right) n_{c}\left( {\bf r}\right) \right) .
\end{eqnarray}
where $K_{3}$ is the three-body collisional cross section and $t$ is the
time duration of one step. For $^{7}Li$, $K_{3}$ is about $2.6\times 10^{-28}%
%TCIMACRO{\unit{cm}}%
%BeginExpansion
\mathop{\rm cm}%
%EndExpansion
^{6}/%
%TCIMACRO{\unit{s}}%
%BeginExpansion
\mathop{\rm s}%
%EndExpansion
$ \cite{moerdijkPRA53}. This additional loss process can be taken into
account during the evaporative stage.

\section{Results and discussion}

\subsection{Evaporating the bosonic atoms}

In the experiments on sympathetic cooling by Schreck {\it et al.} \cite
{schreckPRL87,schreckthesis} a mixture of bosons and fermions is cooled from
2 mK to around 9 $\mu $K, starting with $2\times 10^{8}$ $^{7}$Li bosons and
about $10^{6}$ $^{6}$Li fermions, in an anisotropic trap ($\omega _{r}=2\pi
\times 4050$ Hz, $\omega _{a}=2\pi \times 71.17$ Hz). We have simulated this
process and compared with the experimental data \cite{schreckthesis}. The
suited number of steps can be determined by noting that the time needed for
the system to rethermalize is 200 ms and that the reported cooling process
takes 35 s \cite{schreckthesis}. The number of 200 ms steps{\bf \ }must be
of the order of 100-200. For the evolution of the cut-off radius in time, we
have used the same as in \cite{schreckthesis}. The agreement between theory
and experiment, shown in figure \ref{figure1}, is reasonably good.

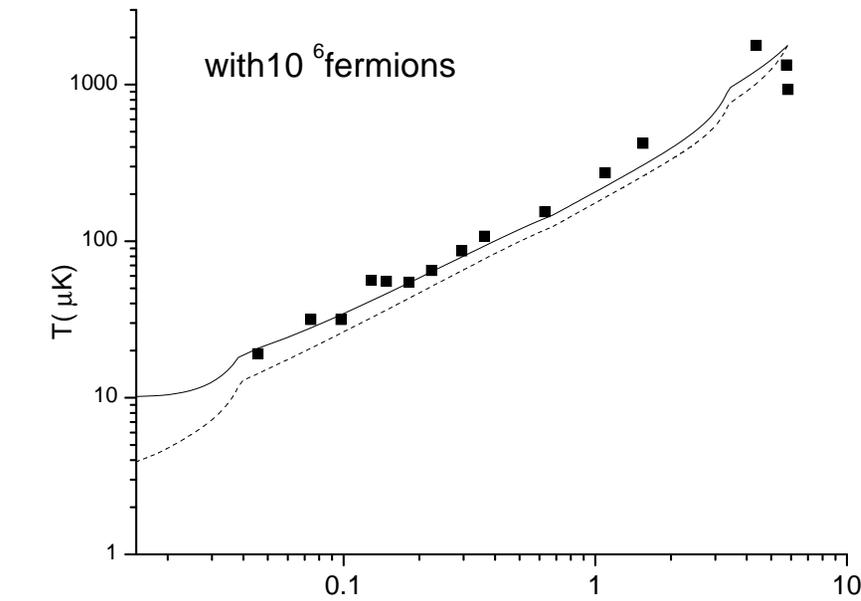
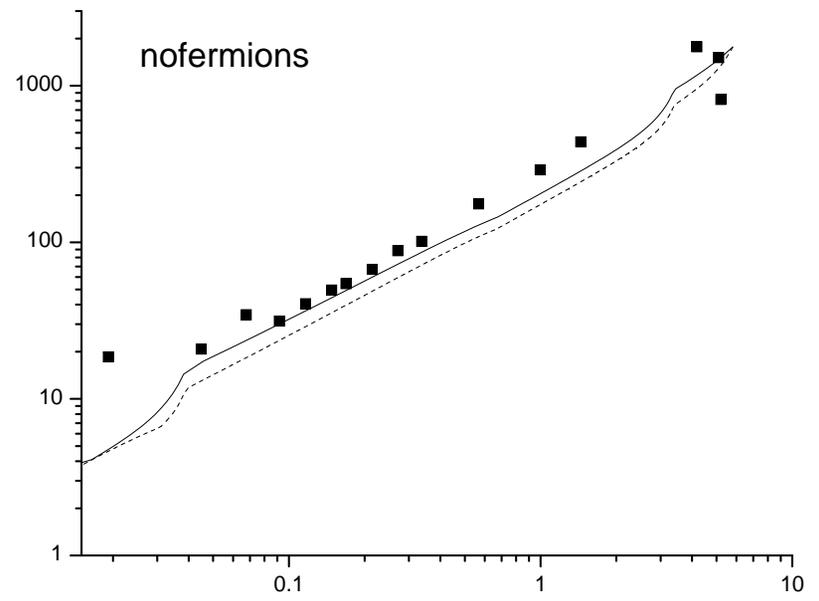
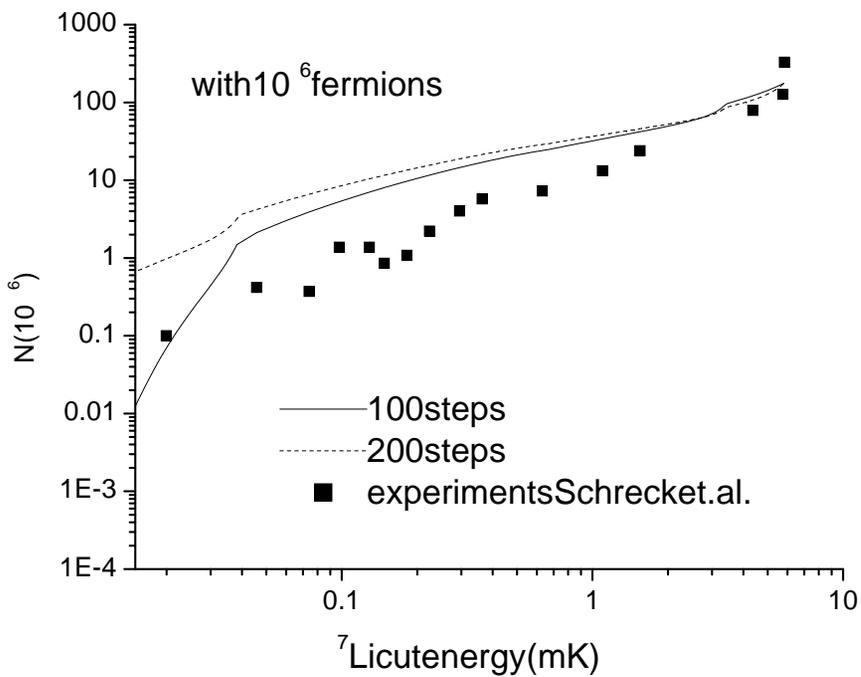
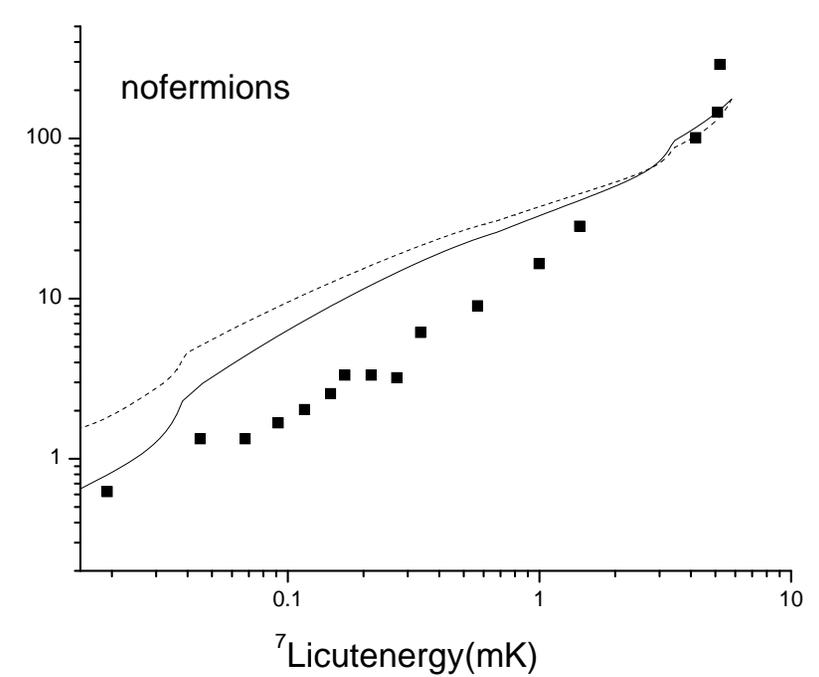
\begin{figure}[tbp]
\caption{The temperature in units of the Fermi temperature is shown for
cooling processes with decreasing cut-off radius, with and without
three-body decay for the condensate. The initial cut-off radius is 20 $%
a_{HO} $ and decreases with 0.32 $a_{HO}$ each time step. The initial
numbers of bosons and fermions are $N_b=10^6$, $N_f=4 \times 10^3$.}
\label{figure2}
\end{figure}

In the next simulation we have started from $9$ $\mu $K, trying to cool
further by evaporating only bosons{\bf ,} as is done in the{\bf \ }%
experiment reported in Ref. \cite{schreckPRL87}{\bf .} We simulate the
evaporative cooling of a system, starting with $N_{f}=4\times 10^{3}$
fermions and $N_{b}=10^{6}$ bosonic atoms at an initial temperature of $T=9$ 
$\mu K$ which is $6.12$ times the Fermi temperature for this amount of
fermions. Figure \ref{figure2} shows the cooling process as a function of
the time (number of evaporation-rethermalization steps), when the
evaporation is performed at decreasing cut-off radius starting from $20$ $%
a_{HO}$ and decreasing with $0.32$ $a_{HO}$ each time step. The lowest
temperature which is obtained is $0.22$ $T_{F}$, in agreement with the
temperature reported by Schreck {\it et al.} \cite{schreckPRL87}. We have
performed the simulation for different values of the three body boson decay
constant ($K_{3}$) and one can see that in our model there is only an effect
on the number of bosons, not on the final temperature.

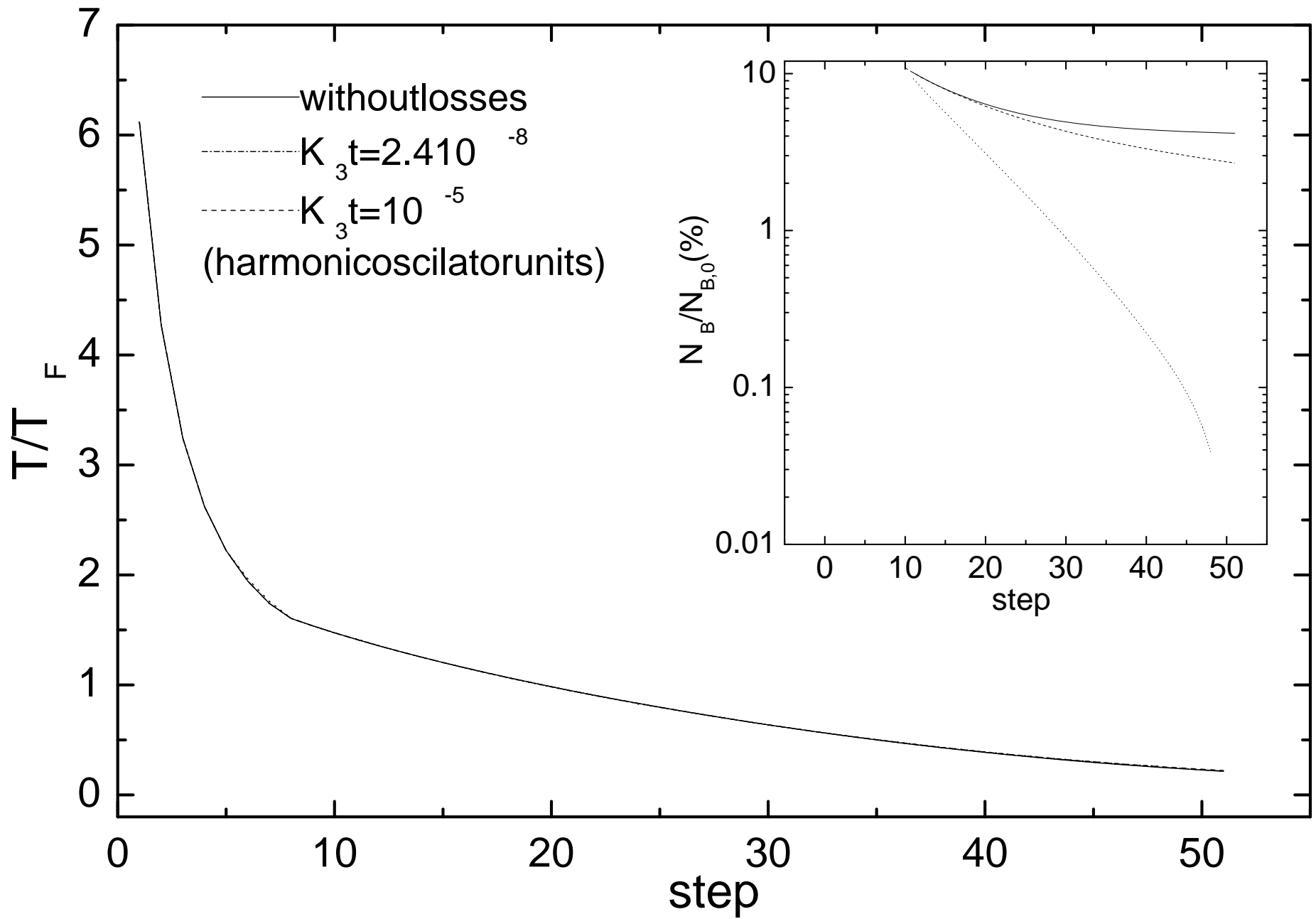
\begin{figure}[tbp]
\caption{ The temperature in units of the Fermi temperature is shown at time
steps 10,20,30,40 and 50 of the cooling process, as a function of the
initial radius $R_{0}$. At each time step, the cut-off radius is reduced by
0.32 $a_{HO}$. The initial numbers of bosons and fermions are $N_b=10^6$, $%
N_f=4 \times 10^3$.}
\label{figure3}
\end{figure}

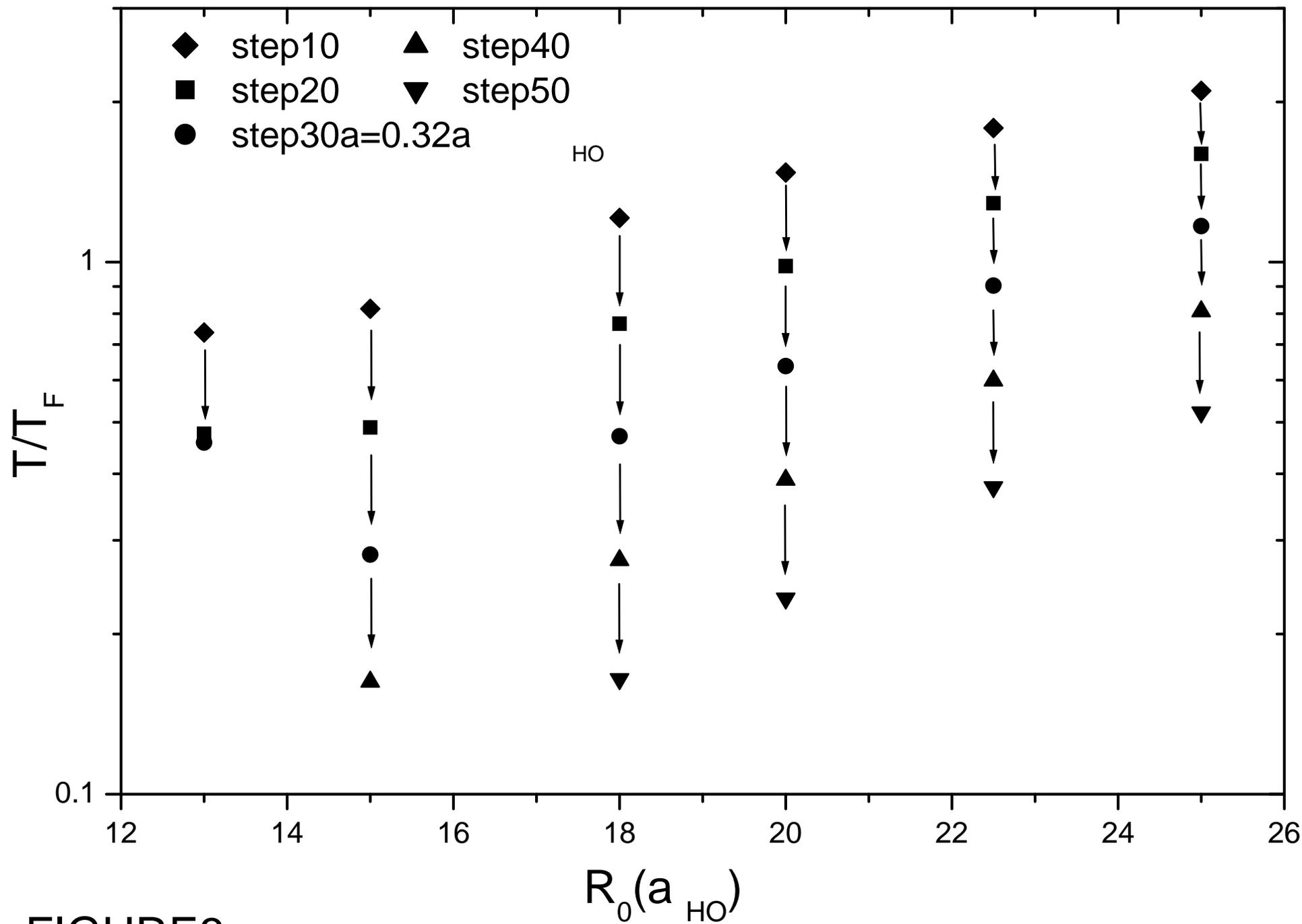
\begin{figure}[tbp]
\caption{The ratio of the temperature to the Fermi temperature is shown as a
function of time (step) in the cooling process. At each step of the cooling
process the cut-off radius is reduced by a value $a$. The different curves
represent differents rates of reduction $a$. The initial radius is always 20 
$a_{HO}.$ The initial numbers of bosons and fermions are $N_b=10^6$, $N_f=4
\times 10^3$.}
\label{figure4}
\end{figure}

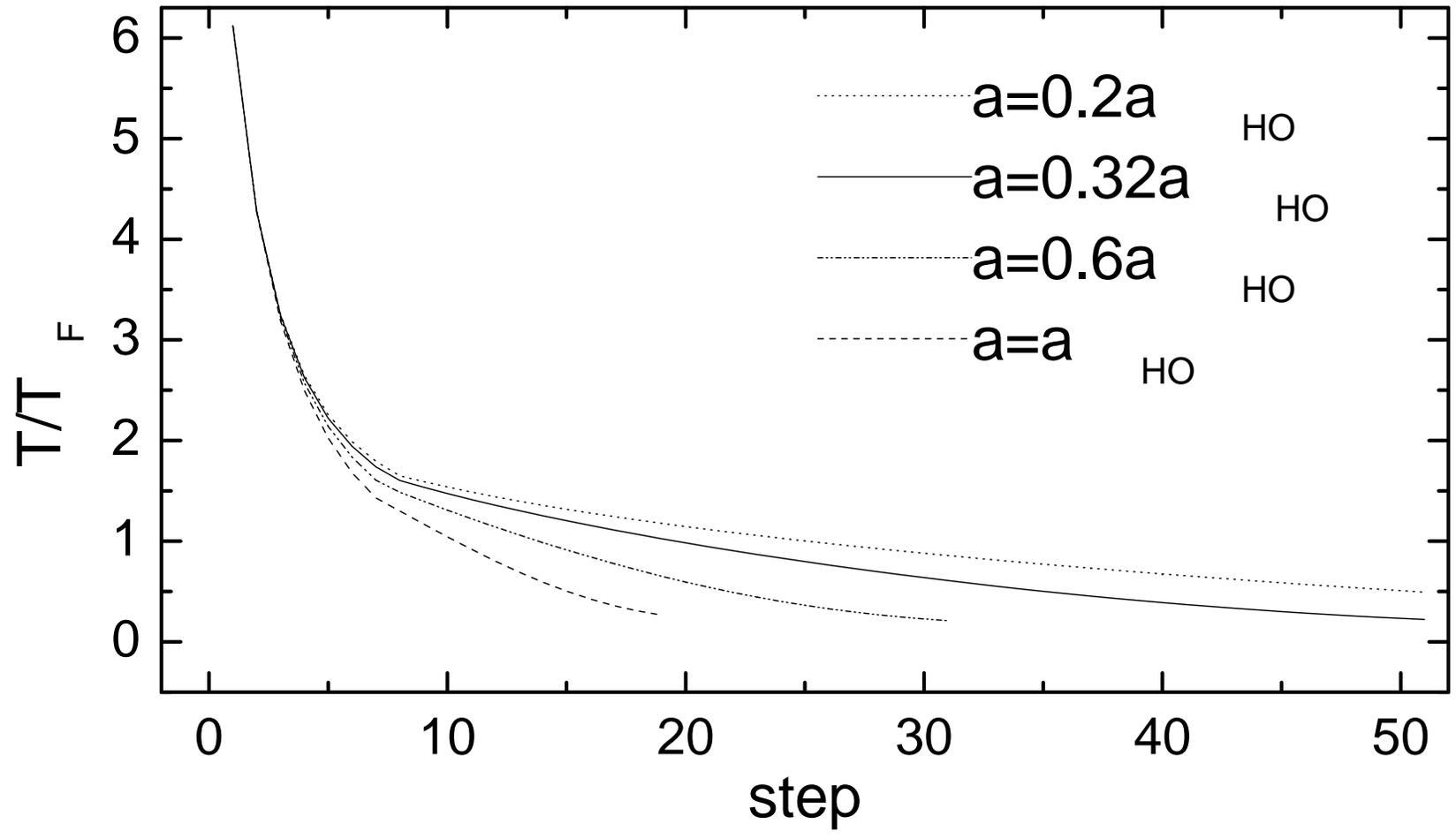
\begin{figure}[tbp]
\caption{The ratio of the temperature to the Fermi temperature for cooling
processes with decreasing radius, for different numbers of initial bosons ($%
N_{B,0}$). De radius starts at 20 $a_{HO}$ and decreases with 0.32 $a_{HO}$
each time step.}
\label{figure5}
\end{figure}

Figure \ref{figure3} illustrates the influence of the initial cut-off radius
on the cooling process, showing the temperature at step 10,20,30,40 and 50
of the cooling process, as a function of the initial cut-off radius. During
each step the cut-off radius was decreased by $0.32$ $a_{HO}$. An{\bf \ }%
optimal radius exists around 18 $a_{HO}$, which is small enough to remove
enough bosons, but sufficiently large in order not to remove too many bosons
in the initial steps and not being energy selective. The simulations have a%
{\bf \ }maximum of{\bf \ }50 steps or stop when the cut-off radius becomes
smaller than the condensate. In figure \ref{figure4} we investigate how the
temperature obtained by cooling is affected by reducing the cut-off{\bf \ }%
radius with different amounts during the cooling process. At each time step,
the cut-off{\bf \ }radius is decreased by a size $a$. The different curves
in the figure show the temperature as a function of time (step in the
cooling process) at different values of $a$. We find that a faster rate of
decrease{\bf \ }of the radius leads to a faster cooling, but that no lower
temperatures can be reached. If the radius is decreased too fast, it reaches
the radius of the condensate too soon, and the cooling process is stopped.
For a fixed time of the cooling process, which is determined by the loss
rate of the trap, an optimal value of $a$ can be determined. For the 50
steps that we took in figure \ref{figure4}, this is $a$ between 0.32 and 0.6
trap{\bf \ }oscillator lengths. Figure \ref{figure5} shows the dependence of
the temperature on the initial number of bosons, keeping the number of
fermions fixed. We find the counterintuitive result that the initial ratio
between bosons and fermions has only a minor effect on the final temperature
that can be reached: it is always around 0.2 $T_{F}$. This can be understood
by noticing that the number of thermal bosons (which are responsible for the
cooling) is almost independent of the total number of bosons for
temperatures below the condensation temperature. ($N_{t}=N_{b}\left( \frac{T%
}{T_{c}}\right) ^{3}\approx T^{3}\zeta \left( 3\right) $).

Truscott {\it et al.} \cite{truscottSCI291} note that the cooling process
seems to stall as soon as the specific heat of the bosons is smaller than
the specific heat of the fermions, because the energy which can be carried
away by evaporating a bosonic atom then becomes smaller than the energy of a
fermionic atom. This result is confirmed by the present simulation, which
explains why the final temperature is insensitive to the initial ratio of
bosons to fermions. Basically, when the number of thermal bosons is small
compared to $N_{F}$, the energy which can be carried away by evaporating the
bosons is small compared to the energy of the Fermi gas. The temperature
does not decrease significantly and the bosons are quickly lost, because the
spatial extent of the cloud remains big.

\subsection{An improved cooling process: sequential evaporation}

This leads us to suggest another cooling process. When evaporative cooling
by removing bosons has led to a final temperature, and continued evaporation
of the bosons no longer leads to substantial further cooling, we suggest to
switch to evaporating the fermions, or fermions and bosons. Removing the
bosons from the trap does not cool the system any more because the bosons
can no longer carry away the necessary energy - but the fermions still can.

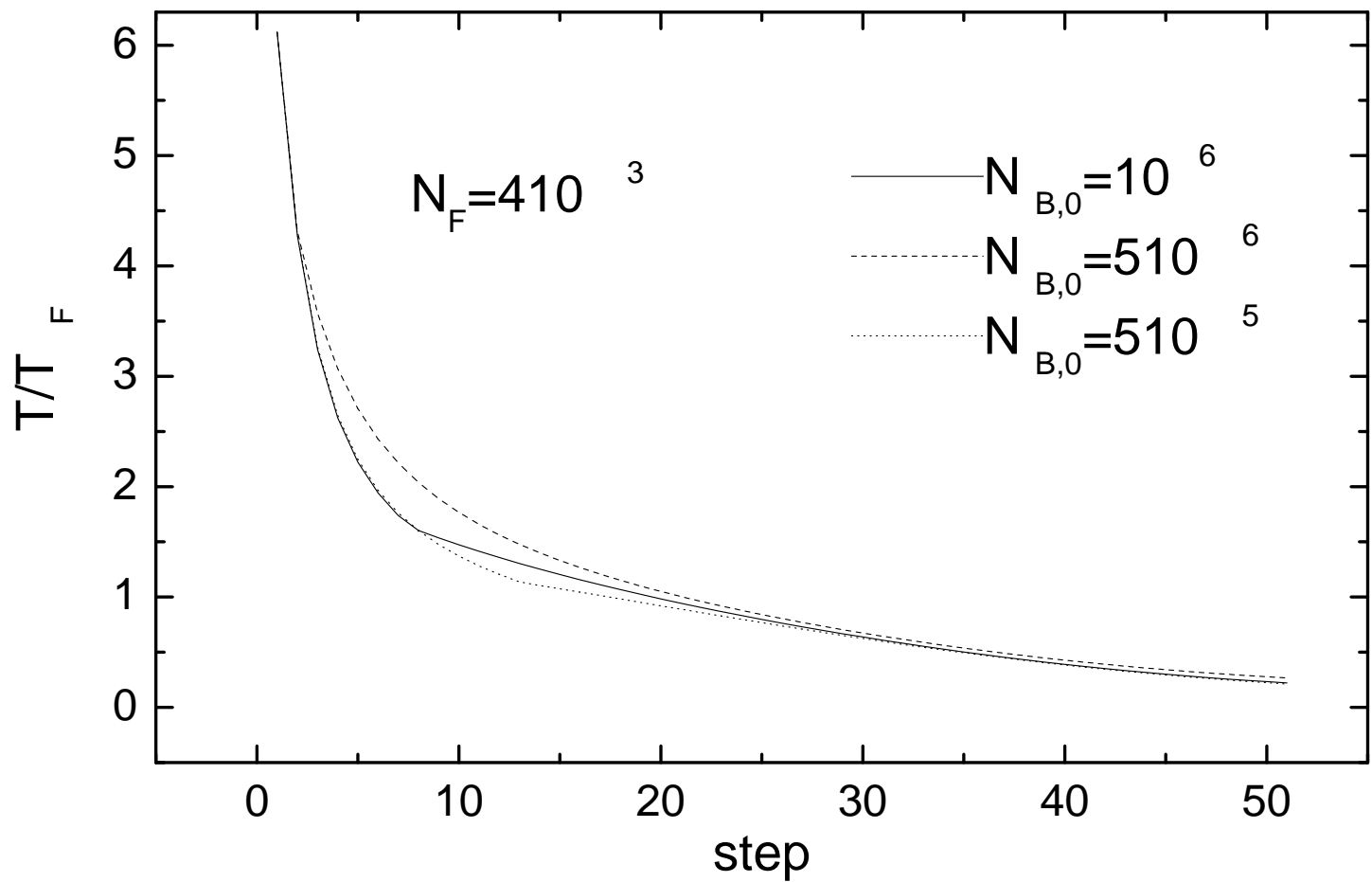
\begin{figure}[tbp]
\caption{Last steps of a sequential sympathetic cooling process. The
temperature relative to the Fermi temperature, $T/T_F$ is shown. For (a)-(c)
the number of bosons is $50.000$, for (d), it is $100.000$. The radius is $%
R=(15-0.1\times step)a_{HO} $ for (a), $R=(15-0.2\times step)a_{HO}$ for(b)
and $R=(18-0.2\times step)a_{HO}$ for (c). For (d) the radius is the same as
in (a). The inset shows the influence of the loss due to three body
collisions involving two bosons and one fermion on $T/T_F$, for the
parameters of the best process (a). The considered values of $K_{3BF}t$ are $%
0$, $\ 10^{-8},$ $10^{-7},$ $5\times 10^{-7}$ and $10^{-6}.$}
\label{figure6}
\end{figure}

Assume that by the evaporation of bosons, the temperature has already
decreased to $0.3$ $T_{F},$ and there are $N_{b}=5\times 10^{4}$ and $%
N_{f}=5\times 10^{4}$ atoms left in the trap. This system is experimentally
achievable by evaporating bosons alone. Figure \ref{figure6} shows the
resulting temperature as a function of time, after switching from
evaporating the bosons to evaporating only the fermions. The cut-off radius
was chosen at $R_{C}=15$ $a_{HO}$ and decreases each time step. In this way
the temperature can be lowered to $0.06$ $T_{F}.$ As also illustrated in the
figure, the number of bosons has almost no influence on the cooling process.

Losses due to three particle collisions involving two bosons and one fermion
is expected to limit the achievable temperature, as pointed out by
Timmermans \cite{timmermansPRL87}. The effect of this type of loss can be
investigated in the present formalism in the same way as the effect of the
three-boson collisions. The results are shown in the inset of figure 6, for
a cooling process starting with $5\times 10^{4}$ bosons and $5\times 10^{4}$
fermions. The different curves correspond to different values of the three
particle recombination constant $K_{3BF}$. If the product of the time to
achieve equilibrium ($t$) and the recombination constant is $%
K_{3BF}t\lessapprox 10^{-7}$ \ we find that the lowest achievable
temperature is raised by less than $0.05$ $T_{F}$. For $K_{3BF}t\lessapprox
10^{-8}$ there is no appreciable effect on the final temperature of the
cooling process.

\bigskip

\section{Conclusions}

We have proposed a versatile, and straightforwardly implementable
three-fluid model to describe the sympathetic cooling of bosons and
fermions. Our model reproduces well the cooling observed in an experiment
with $^{6}$Li and $^{7}$Li, and we find that the final temperature $%
T/T_{F}\approx 0.2$ which is reached is not sensitive to the initial ratio
of the number of bosons to the number of fermions. Lower temperatures can be
reached if the stage where the only the bosonic atoms are evaporated is
followed by a stage where also the fermionic atoms are evaporated. When this
procedure is applied, a temperature of $T/T_{F}=0.06$ could be reached,
which in combination with the tunability of the interaction strength through
the Feshbach resonance, brings the Fermi gas closer to temperatures where
pairing may be observed.

\section{Acknowledgments}

Two of the authors (M. W. and J. T.) are supported financially by the Fund
for Scientific Research - Flanders (Fonds voor Wetenschappelijk Onderzoek --
Vlaanderen). This research has been supported by the\ GOA BOF UA 2000, IUAP,
the FWO-V projects Nos. G.0071.98, G.0306.00, G.0274.01, WOG WO.025.99N
(Belgium).

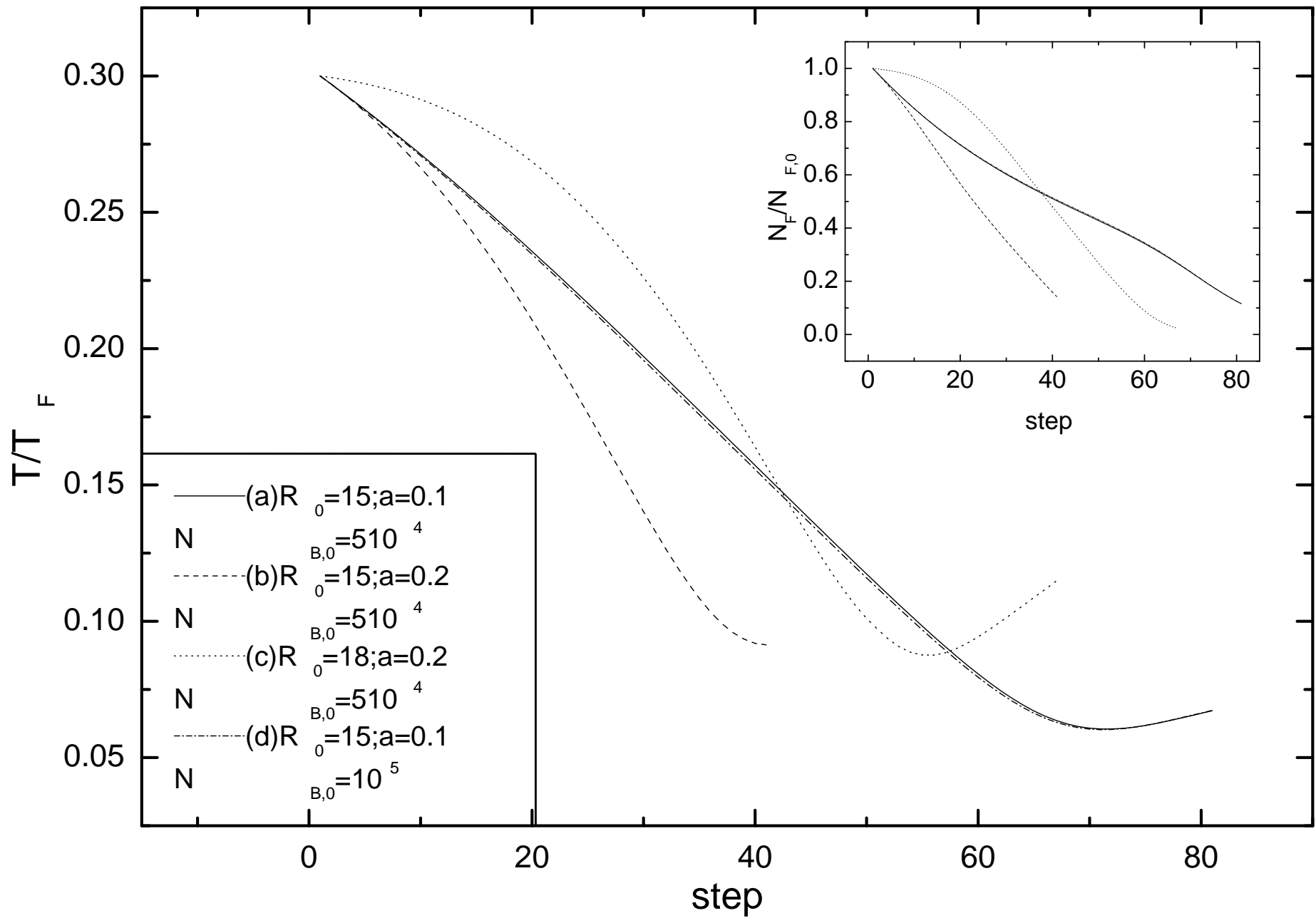
\end{document}